\documentclass[letterpaper,12pt]{article}
\usepackage{tabularx} 
\usepackage{amsmath}  
\usepackage{amssymb}
\usepackage{graphicx} 
\usepackage{subfig}
\usepackage{natbib}
\usepackage{dcolumn}
\usepackage[margin=1in,letterpaper]{geometry} 
\usepackage{cite} 
\usepackage[final]{hyperref} 
\hypersetup{
	colorlinks=true,       
	linkcolor=blue,        
	citecolor=blue,        
	filecolor=magenta,     
	urlcolor=blue         
}
\usepackage{float}

\begin{document}
\newcommand{\ou}{%
  \mathrel{%
    \vcenter{\offinterlineskip
      \ialign{##\cr$+$\cr\noalign{\kern-1.5pt}$-$\cr}%
    }%
  }%
}
\title{Internal Technical Report: uGMRT Band 4 Polarimetry}

\begin{figure}
\centering{
\includegraphics[width=5cm]{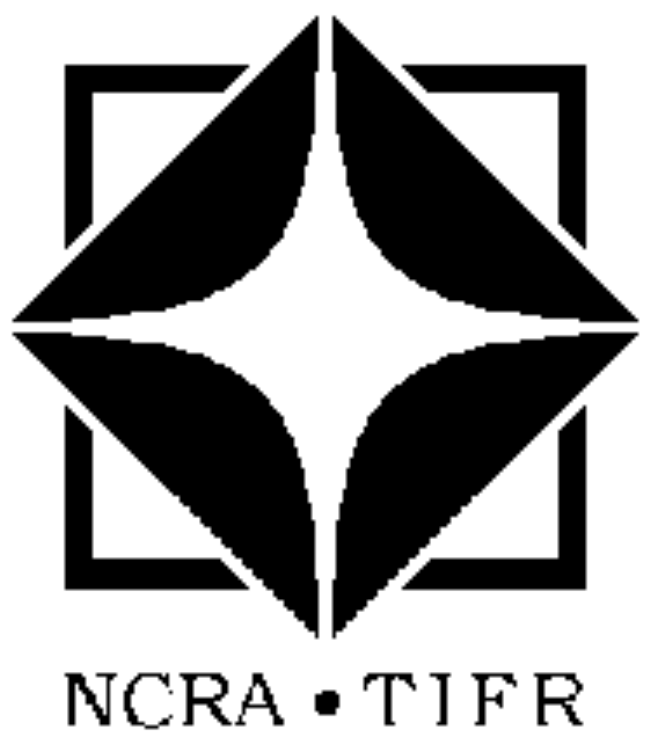}}
\end{figure}

\author{Preeti Kharb, Silpa Sasikumar, Janhavi Baghel, Salmoli Ghosh}
\date{\today}
\maketitle

\begin{abstract}
This is a technical report for band 4 ($550-900$~MHz) polarization data with the upgraded GMRT (uGMRT). The report  describes the band 4 polarization data analysis procedure and includes notes for observers who are planning polarization observations with the uGMRT. A few pipelines that are currently being used and tested by astronomers at NCRA are discussed as well.
\end{abstract}
\section{Introduction}
The upgraded GMRT (uGMRT) operates in five observing bands, namely, band 2 ($120-250$~MHz), band 3 ($250-500$~MHz), band 4 ($550-900$~MHz) and band 5 ($1000-1450$~MHz) \citep{Swarup1991,Gupta2017}. These feeds are mounted on a rotating turret at the focus of the reflecting surface, which is a parabolic wiremesh. The front-end of lower frequency bands (band 2 - band 4) employ dipole antenna feeds in `crossed' or `boxing ring' configuration to receive dual linear polarizations (vertical and horizontal polarization). These are converted to circular polarizations (right hand circular polarization, RCP, and left hand circular polarization, LCP) using a polarizer placed before the low noise amplifier (LNA). The band 5 front-end consists of a corrugated horn feed along with an orthomode transducer (OMT). The OMT picks up two linear orthogonal components (vertical and horizontal polarization) of the incoming signal and feeds them directly to the LNA. No polarizer is placed in the signal path in order to maintain the low system temperature. Circularly polarized signals are received in all the observing bands except at band 5, which produces linearly polarized signals. The orthogonal polarizations from each antenna (R and L or X and Y) are brought to the control room for further processing. See more details about the GMRT antennas and feeds in the Low Frequency Radio Astronomy (Blue) book, {\url{http://www.ncra.tifr.res.in/ncra/gmrt/gmrt-users/low-frequency-radio-astronomy}}. 

\section{Calibrating Polarization Data from Band 4}
In this section, we discuss the steps to be followed for carrying out polarization calibration at band 4 with the uGMRT. For details related to polarization concepts, the reader is referred to the Synthesis Imaging in Radio Astronomy II book \citep{Synthesis1999}. Earlier polarization observations from the GMRT have been presented by \citet{Joshi2011} as well as \citet{Farnes2013}. We describe below the strategy adopted for band 4 polarization data analysis with the uGMRT, results from which have been presented in \citet{Silpa2021} and \citet{Baghel2022}. 

The uGMRT band 4 polarization data analysis follows the standard steps of carrying out the leakage calibration using either an unpolarized calibrator or a well-studied polarized calibrator, followed by polarization angle calibration using a well-studied polarized calibrator with a known polarization position angle. These polarization calibration steps are performed after the basic gain calibration which is well documented for uGMRT band 4 data \citep[e.g.,][]{Ishwara2020,Silpa2020}. A Python script has been developed by us for carrying out the initial gain calibration followed by polarization calibration using $\tt{CASA}$ and is available at {\url{https://sites.google.com/view/silpasasikumar/}}. This pipeline builds on the one developed by C. H. Ishwara-Chandra et al.\footnote{\url{http://www.ncra.tifr.res.in/~ishwar/pipeline.html}} and  includes modifications to deal with polarization data. A semi-automated version with polarization models for the standard calibrators included, can be accessed from here:\\ {\url{https://github.com/jbaghel/Improved-uGMRT-polarization-pipeline}}. Currently, all these pipelines have been tested extensively on band 4 data, although a few tests have been carried out also on band 3 data. A {\tt CAPTURE}-based version \citep{Kale2021} of the polarization pipeline, {\tt CAPTURE-POL}, is available here:\\ {\url{http://www.ncra.tifr.res.in/~ruta/IDAP/index.html}}. 

The calibration steps are described in detail now. A detailed explanation of the steps can also be found in the EVLA polarization tutorial\footnote{\url{https://casaguides.nrao.edu/index.php?title=VLA\_Continuum\_Tutorial\_3C391-CASA4.6}}. The pipeline flags data from all four correlations (RR, LL, RL, LR) individually, with special care taken for the polarization calibrator data. The pipeline then (i) solves for cross-hand (RL, LR) delays, which result from residual delay difference between R and L signals. This is solved using a polarized calibrator which has strong cross$-$polarization (RL, LR). The task $\tt{GAINCAL}$ in $\tt{CASA}$ with $\tt{gaintype=KCROSS}$ is used for this. The pipeline then (ii) solves for instrumental polarization (i.e, the frequency-dependent leakage terms or `D-terms') using either an unpolarized calibrator or a polarized calibrator with a good parallactic angle coverage. This is necessary to correct for the polarization leakage between the feeds owing to their imperfections and non-orthogonality (e.g. the R/X-polarized feed picks up L/Y-polarized emission, and vice versa). The task $\tt{POLCAL}$ in $\tt{CASA}$ with $\tt{poltype=Df}$ for the unpolarised calibrator and $\tt{poltype=Df+QU}$ for the polarized calibrator is used. This task uses the Stokes I, Q, and U values (Q and U being zero for an unpolarized calibrator) in the model data to derive the leakage solutions. Also, when $\tt{poltype=Df+QU}$ is used, the task solves for both source polarization and instrumental polarization simultaneously. This is based on the principle that the source polarization angle rotates with parallactic angle whereas the instrumental polarization angle does not. 

Finally, the pipeline (iii) solves for frequency-dependent polarization position angle. Polarization position angle for circular feeds is unknown $\it{a priori}$, therefore is accurately determined using a polarized source with known polarization electric vector position angle (EVPA). A polarization model is set for the polarized calibrator using the task $\tt{SETJY}$ in $\tt{CASA}$, then solve for the angle using the task $\tt{POLCAL}$ with $\tt{poltype= Xf}$. Basic and polarization calibration is followed by the task $\tt{FLUXSCALE}$ in $\tt{CASA}$ to determine flux density values of the phase and polarization calibrators. All the calibration solutions are then applied to the multi-source target dataset. The $\tt{CASA}$ task $\tt{SPLIT}$ is used to create visibility subsets for individual targets from the original dataset while averaging the spectral channels such that the bandwidth smearing effects are negligible. 

Each target data is finally imaged for total intensity or Stokes `I' using the Wide-Band, Wide-Field Imaging algorithm of $\tt{TCLEAN}$ task in $\tt{CASA}$. Several iterations of phase-only self-calibration and minimum two iterations of amplitude and phase self-calibration are performed for Stokes `I' images until convergence. The parameter $\tt{PARANG}$ in task $\tt{APPLYCAL}$ during the self-calibration cycles, is set to $\tt{FALSE}$. Stokes `Q' and `U' images are obtained using the last self-calibrated dataset and the same input parameters as for Stokes `I' images. Linear polarization (P=$\sqrt{(Q^2+U^2)}$) and polarization angle ($\chi$=tan$^{-1} (Q/U)$) images are obtained by combining the Stokes `Q' and `U' images using the task $\tt{COMB}$ in $\tt{CASA}$ with $\tt{opcode=POLC}$ and $\tt{POLA}$ respectively. $\tt{POLC}$ takes care of the Ricean bias correction. Regions with signal-to-noise ratio less than three times the {\it rms} noise in the linear polarization image and with values greater than 10$^{\circ}$ error in the polarization angle image are banked while using $\tt{COMB}$. Fractional polarization image F=P/I is obtained by dividing the total intensity image by the linear polarization image using the task $\tt{COMB}$ in $\tt{CASA}$ with $\tt{opcode=DIV}$. Regions with values greater than 10\% error in the fractional polarization image are blanked. 

\section{Choosing Polarization Calibrators at Band 4}
Commonly used calibrators for polarization calibration are 3C286 and 3C138, which are polarized sources at band 4, and 3C84 and OQ208, which are unpolarized sources at band 4. The two other polarized calibrators often used in GHz frequency observations,\footnote{\url{https://science.nrao.edu/facilities/vla/docs/manuals/obsguide/modes/pol}} viz., 3C48 and 3C147 are typically unpolarised in band 4, but may work for band 5 of the uGMRT. Note that 3C147 may work for leakage calibration but not polarization angle calibration. Band 4 polarization experiments have indicated that OQ208 is not a good leakage calibrator for uGMRT since it produces unphysical `D-term' values. It is a relatively faint source ($<1$~Jy at $\sim1$~GHz). Therefore, a single short scan does not provide sufficient signal-to-noise ratio to accurately determine the instrumental polarization. Recent experiments have suggested that apart from 3C147 or 3C48, the source J0713+4349, which is unpolarized in band 4 (see the Category C leakage calibrators in $-$\\ {\url{https://science.nrao.edu/facilities/vla/docs/manuals/obsguide/modes/pol}}), also works well for band 4 leakage calibration. One or two scans of these unpolarized calibrators are recommended for polarization experiments. For good leakage calibration, it is recommended that the polarized calibrators, 3C286 or 3C138, be observed in at least 4 scans with a parallactic angle coverage that is close to $60^\circ$, and that these scans are spread evenly across the experiment. 

{\it We note that the polarization analysis for band 3 data is identical to band 4 data analysis. However, the standard polarization calibrator, 3C286, is too weakly polarized at band 3 ($\le3\%$) to work as a calibrator for polarization angle. In this case, an unpolarized calibrator like 3C147 can be used for leakage calibration, and the eastern polarized hotspot in DA240 can be used for the polarization angle calibrator (it has a fractional polarization of $\sim$20\% in band 3 and a polarization angle close to $\sim +68.5^{\circ}$; see the EVLA Memo 207\footnote{{\url{https://library.nrao.edu/public/memos/evla/EVLAM_207.pdf}}} \citep{perley2019evla}. We are currently in the process of analysing several more polarization datasets at band 3, the results from which will be presented in a future report.}

\section{Note for Observers$^\dagger$}
{\bf $^\dagger$See also {\url{http://gmrt.ncra.tifr.res.in/gmrt\_users/recent\_updates.html}}}\\

We discuss below the main points to be taken into consideration when planning polarimetric observations with the uGMRT. Documentation is also available at $-$\\
{\url{http://www.gmrt.ncra.tifr.res.in/doc/polarisation_observation_sop.pdf}} \\

While preparing the proposal and using the GMRT exposure time calculator (ETC) \\
({\url{http://www.ncra.tifr.res.in:8081/\~secr-ops/etc/rms/rms.html}}) 

\begin{enumerate}
\item For Point 7, choose “Number of polarizations = 2”,

\item For Point 17, under “Extra Bandpass/Polarization Time” choose additional time for multiple scans of the polarization calibrator for good parallactic angle coverage as well as slewing time. Typically, the total “overhead” time for a polarization experiment can be $\sim50$\%.
\end{enumerate}

During proposal preparation on GMRT NAPS ({\url{https://naps.ncra.tifr.res.in/naps/login}})
\begin{enumerate}
\item Under the “Observation Details” tab, choose the option “GWB Interferometer Polar”,

\item For “Special Requirements”, add “Visibility of a polarized calibrator (3C286, 3C138)** and/or an unpolarized calibrator (3C84, 3C147, J0713+4349)** throughout each observing block is critical for robust polarization calibration. The polarized calibrator will need to be observed in several short scans for good parallactic angle coverage.”
\end{enumerate}

While creating the command files for the observations \\
({\url{http://www.ncra.tifr.res.in/\~secr-ops/cmd/cmd.html}})

\begin{enumerate}
\item Choose “Stokes Parameter: Full\_Polar(4)”,

\item Under “Special requirement or additional info (if any):”, add “Visibility of a polarized calibrator (3C286, 3C138)** and/or an unpolarized calibrator (3C84, 3C147, J0713+4349)** throughout each observing block is critical for robust polarization calibration. The polarized calibrator will need to be observed in several short scans for good parallactic angle coverage.”

\item If you chose the secondary backend configuration as GSB during proposal preparation, then add under “Special requirement” that “Stokes Parameter: Full\_Polar(4) must be used for GSB as well.”

\item **Pick one polarized/unpolarized calibrator from the list above that works for your experiment. For additional calibrators you can consult $-$\\
{\url{https://science.nrao.edu/facilities/vla/docs/manuals/obsguide/modes/pol}}. Note however that the testing of these calibrators for uGMRT data is still ongoing. {\it OQ208 does not work well as an unpolarised calibrator for $<1~$GHz observations with the uGMRT due to its low flux density.} 
\end{enumerate}

We note a couple of points to be taken into consideration when planning polarimetric observations with the uGMRT. 

\begin{enumerate}

\item While preparing the proposal and using the GMRT exposure time calculator\footnote{\url{http://www.ncra.tifr.res.in:8081/\~secr-ops/etc/rms/rms.html}} make sure that ``Number of polarizations = 2" for Point 7, and for ``Extra Bandpass/Polarization Time'' under Point 17, include additional time for multiple scans of the polarization calibrator for a good parallactic angle coverage as well as slewing time. Typically, the total ``overhead'' for a polarization experiment can be $\sim50$\%. 

\item During the proposal preparation on GMRT NAPS website\footnote{\url{https://naps.ncra.tifr.res.in/naps/login}}, choose the option ``Choose GWB Interferometer Polar'' and under ``Special requiremnent'' add ``Visibility of an unpolarized (OQ208 or 3C84) and/or a polarized calibrator (3C286 or 3C138) throughout each observing block is critical for robust polarization calibration. The polarized calibrator will need to be observed in several short scans for good parallactic angle coverage."
\end{enumerate}

If you have been awarded uGMRT observing time, note the following while creating the command files\footnote{\url{http://www.ncra.tifr.res.in/\~secr-ops/cmd/cmd.html}}. 

\begin{enumerate}
\item Choose ``Stokes Parameter: Full\_Polar(4)'', and under ``Special requirement or additional info (if any): '' add ``Visibility of one unpolarized (OQ208 or 3C84) and one polarized calibrator (3C286 or 3C138) throughout each observing block is critical for robust polarization calibration. The polarized calibrator will need to be observed in several short scans for good parallactic angle coverage.''. 

\item During the analysis of data, make sure to flag bad channels in all antennas if those channels have been flagged for the reference antenna for the leakage calibrator. Otherwise, the `D-term' amplitudes tend to blow up (become $>100\%$) for those channels.
\end{enumerate}

\section{Polarization Errors at Band 4}
Errors in polarization experiments comprise of errors in the fractional polarization ($\equiv P/I$) and polarization position angles. As described by \citet{Hales2017}, the systematic error in the fractional polarization can be taken as the characteristic `D-term' modulus error, which under Rayleigh statistics is given by $\sigma_{d}=\sqrt{(\pi/2)}~\sigma$, where $\sigma$ is the characteristic error in either the real or imaginary `D-term' values \citep[see][]{Hales2017}. The systematic error in the polarization angle is given by $\phi / \sqrt{N}$, where $\phi$ is the characteristic uncertainty in the mechanical feed alignment per antenna and N is the number of antennas in the array. For the circular feed basis, the spurious level of fractional polarization is given by $\sigma_{d}~\sqrt{(\pi/2~N)}$. Only the fractional polarization values detected above this level can be considered believable in polarization experiments. 

The leakage tables can be directly viewed using the task $\tt{browsetable}$ in $\tt{CASA}$ where the rows represent antennas. The CPARAM column gives the real and imaginary components of the `D-term'  values separately for the RCP and the LCP components per antenna per channel. The $\tt{AIPS}$ task $\tt{PCAL}$ provides the errors in the `D-term' amplitudes and phases for both RCP and LCP components per channel per antenna ($\tt{CASA}$  provides only amplitude errors). Both $\tt{CASA}$ and $\tt{AIPS}$  suggest $\sigma$ to be $\sim1\%$ for the uGMRT antennas. This yields $\sigma_{d}$ to be $\sim1.2\%$, and the spurious fractional polarization level to be $\sim0.3\%$. Finally, additional errors are introduced in the imaging process itself. Average error values can be obtained from the respective noise images that are created while running the $\tt{CASA}$ or $\tt{AIPS}$ task $\tt{COMB}$. Adding these in quadrature to the systematic errors obtained above yields a total fractional polarization $\le 1.5\%$ in band 4 for a source flux density of $\sim$10 mJy \citep[e.g.,][]{Silpa2021}. Obtaining the error in the polarization angle requires $\phi$ which is the characteristic uncertainty in feed alignment per antenna. This is yet to be determined for the uGMRT. For ALMA, a $\phi=2^\circ$ for an antenna results in the  polarization angle error of $0.3^\circ$ for the array with 40 antennas. 

\subsection{Ionospheric Corrections}
While ionospheric effects are relevant to all of the uGMRT bands because of the low radio frequencies, ionospheric corrections have not been carried out for the polarization data obtained thus far. This is because there are currently no widely available models to correct for differences in the observing geometry or the varying magnetic field strength of the ionosphere over the uGMRT antennas. Using the total electron content (TEC) maps alone, \citet{Farnes2014} had noted that the maximum Faraday rotation due to the ionosphere was $\leq2$~rad~m$^{-2}$ at 610~MHz for the legacy GMRT for the duration of their observations ($3-4$ hours on source). More refined estimates for the uGMRT still need to be carried out. We note that ionospheric corrections of the order of $0.1-0.3$~rad~m$^{-2}$ have been noted for LOFAR observations at 150~MHz by \citet{Mahatma2021}. 

\section{Antenna Leakages at Band 4}
The average instrumental leakage amplitudes for most GMRT antennas is between $5-10$\% at band 4. Leakage amplitudes have been found to be typically less than 15 to 20\%. However, for some antennas and some (typically end) channels the amplitudes can go as high as $35-40\%$ (for several experiments in 2022, antennas C06, C09, W03 and S06 were the ones showing higher leakage amplitudes). A comparison of the `D-term' amplitude variation with frequency or channels of antennas across different experiments shows that they are consistent in shape but may differ in their maximum amplitudes. Overall the `D-term' amplitude patterns are similar between experiments for a given antenna. In figures 1 to 9, we present the leakage plots obtained for a couple of experiments using either several scans of a polarized calibrator like 3C286 or one or two scans of an unpolarized calibrator like 3C147 or  J0713+4349. The blue points denote the RCP signal while the green points denote the LCP signal.


\begin{figure}
\includegraphics[width=19cm,trim=185 40 20 20]{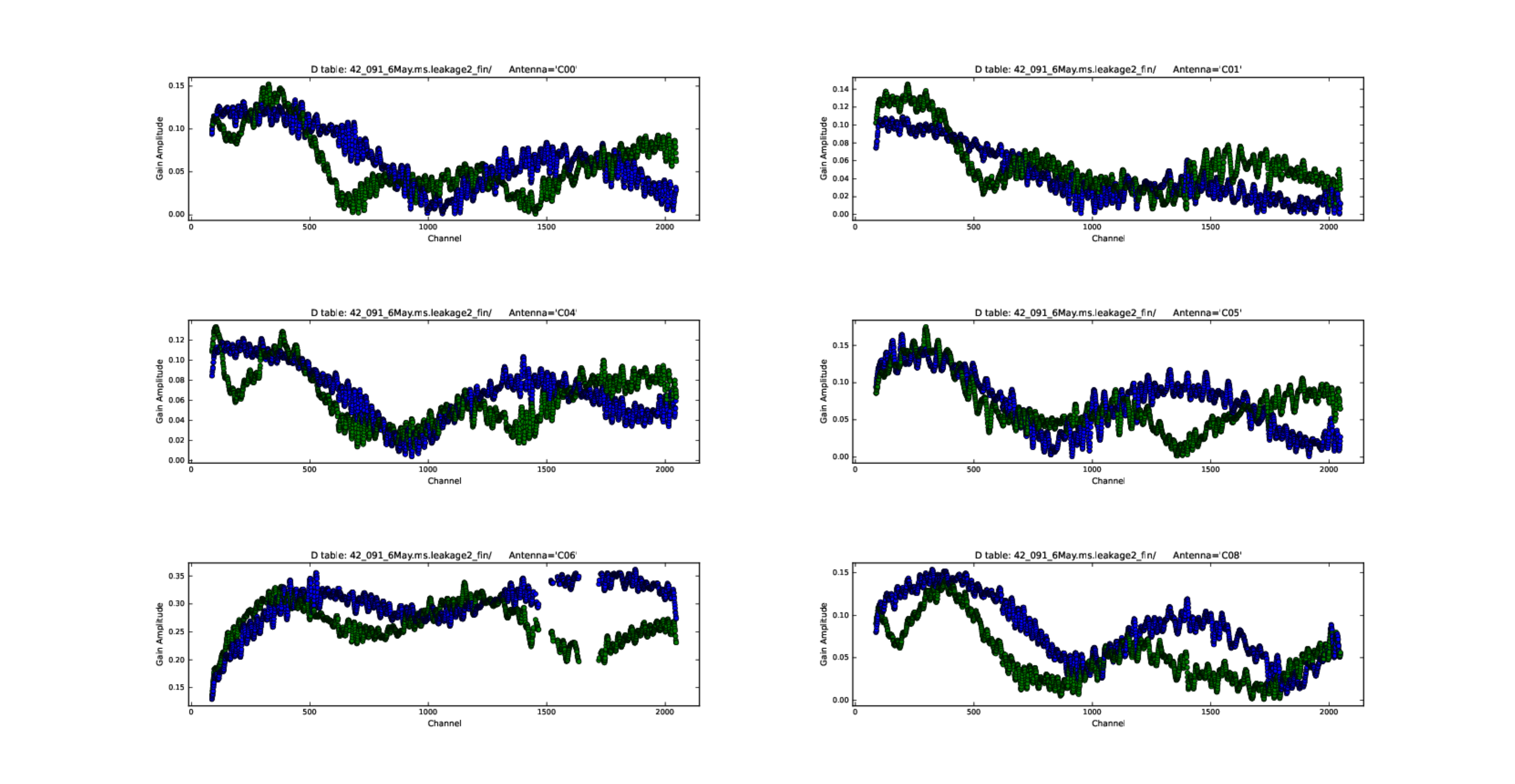}
\includegraphics[width=19cm,trim=185 40 20 20]{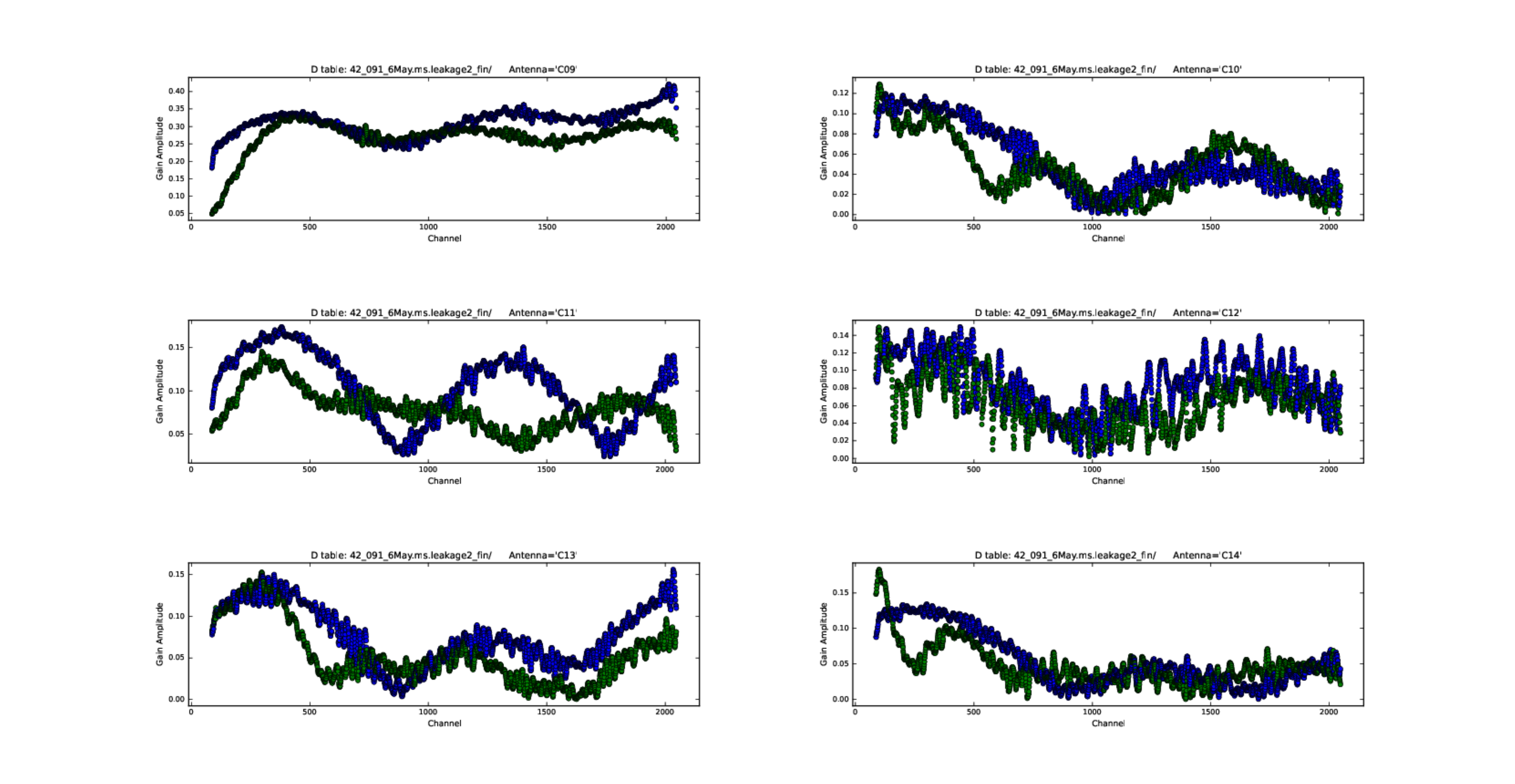}
\caption{\small Antenna `D-term' amplitudes vs channels for the Project 42\_091 from 6 May 2022 using the unpolarized calibrator 3C147.}
\end{figure}
\begin{figure}
\includegraphics[width=19cm,trim=185 40 20 20]{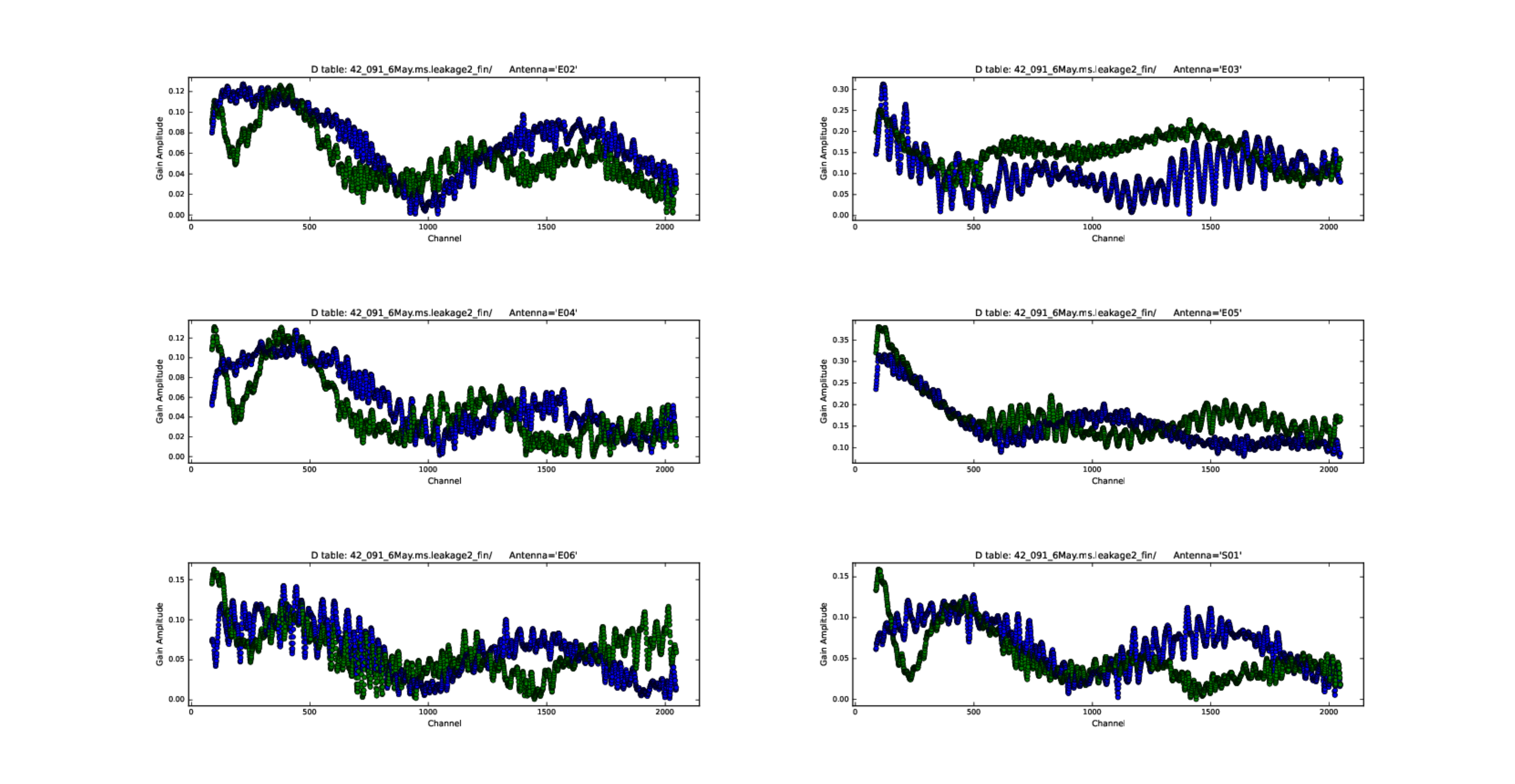}
\includegraphics[width=19cm,trim=185 40 20 20]{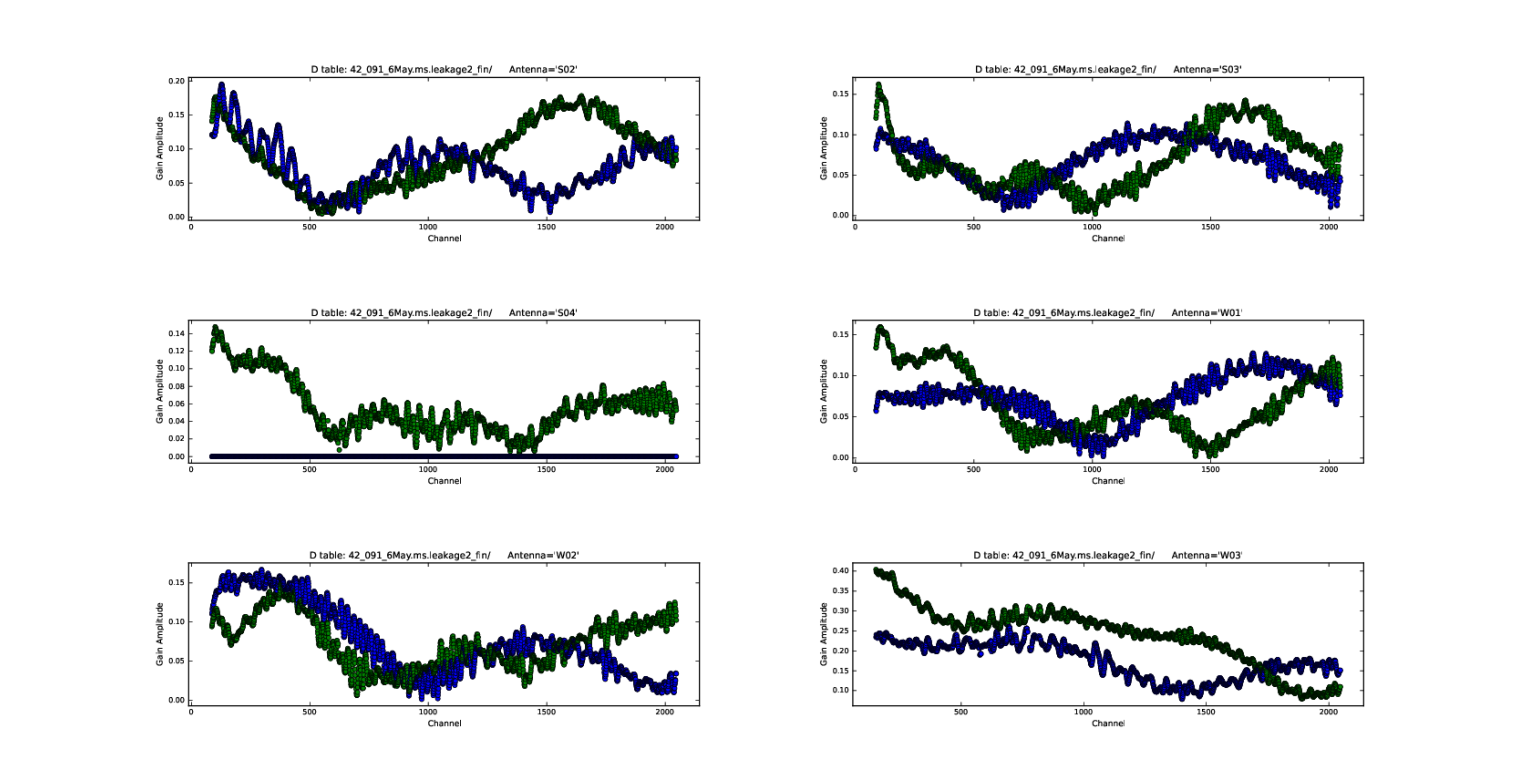}
\caption{\small Antenna `D-term' amplitudes vs channels for the Project 42\_091 from 6 May 2022 using the unpolarized calibrator 3C147.}
\end{figure}
\begin{figure}
\includegraphics[width=19cm,trim=185 40 20 20]{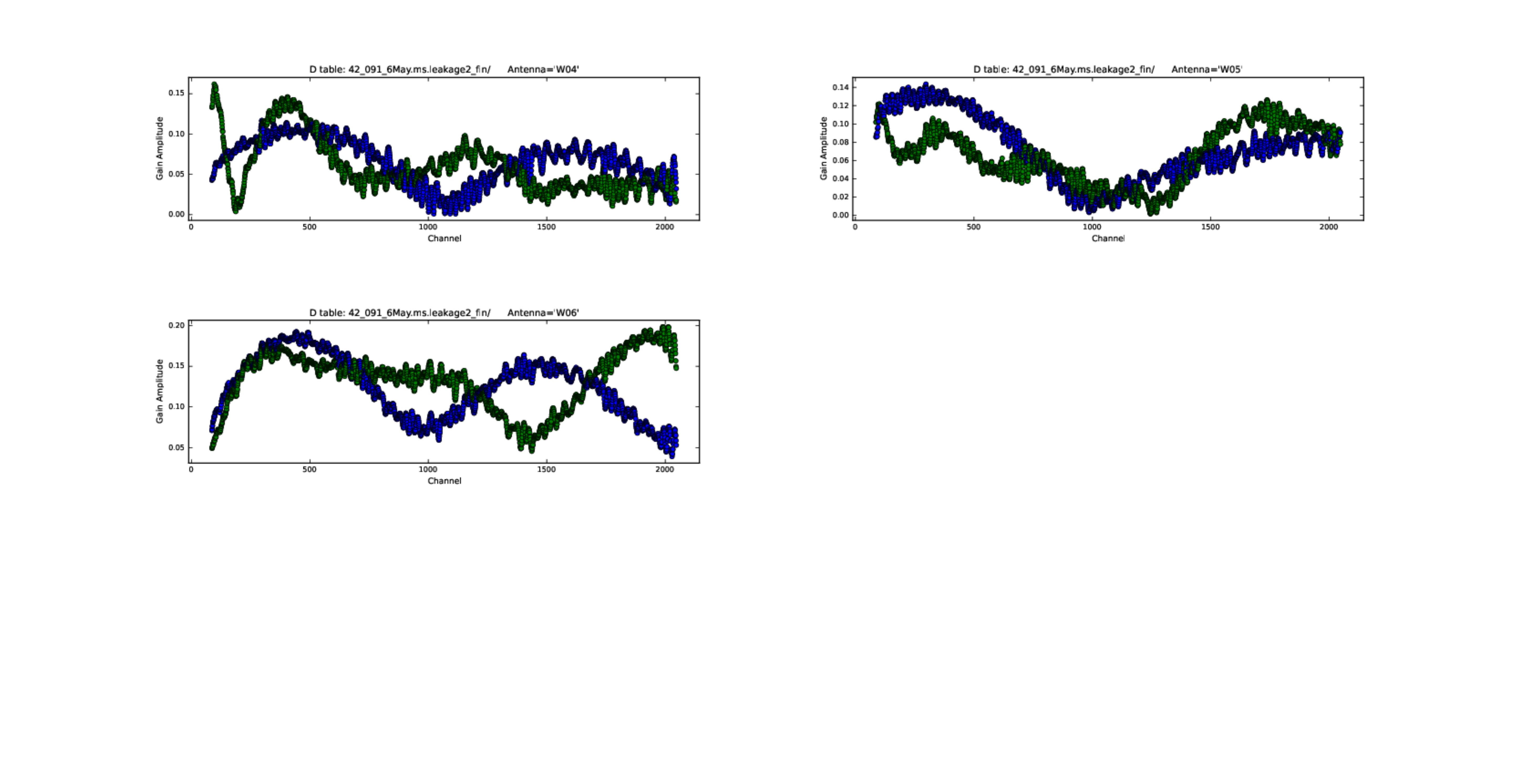}
\caption{\small Antenna `D-term' amplitudes vs channels for the Project 42\_091 from 6 May 2022 using the unpolarized calibrator 3C147.}
\end{figure}

\begin{figure}
\includegraphics[width=19cm,trim=185 40 20 20]{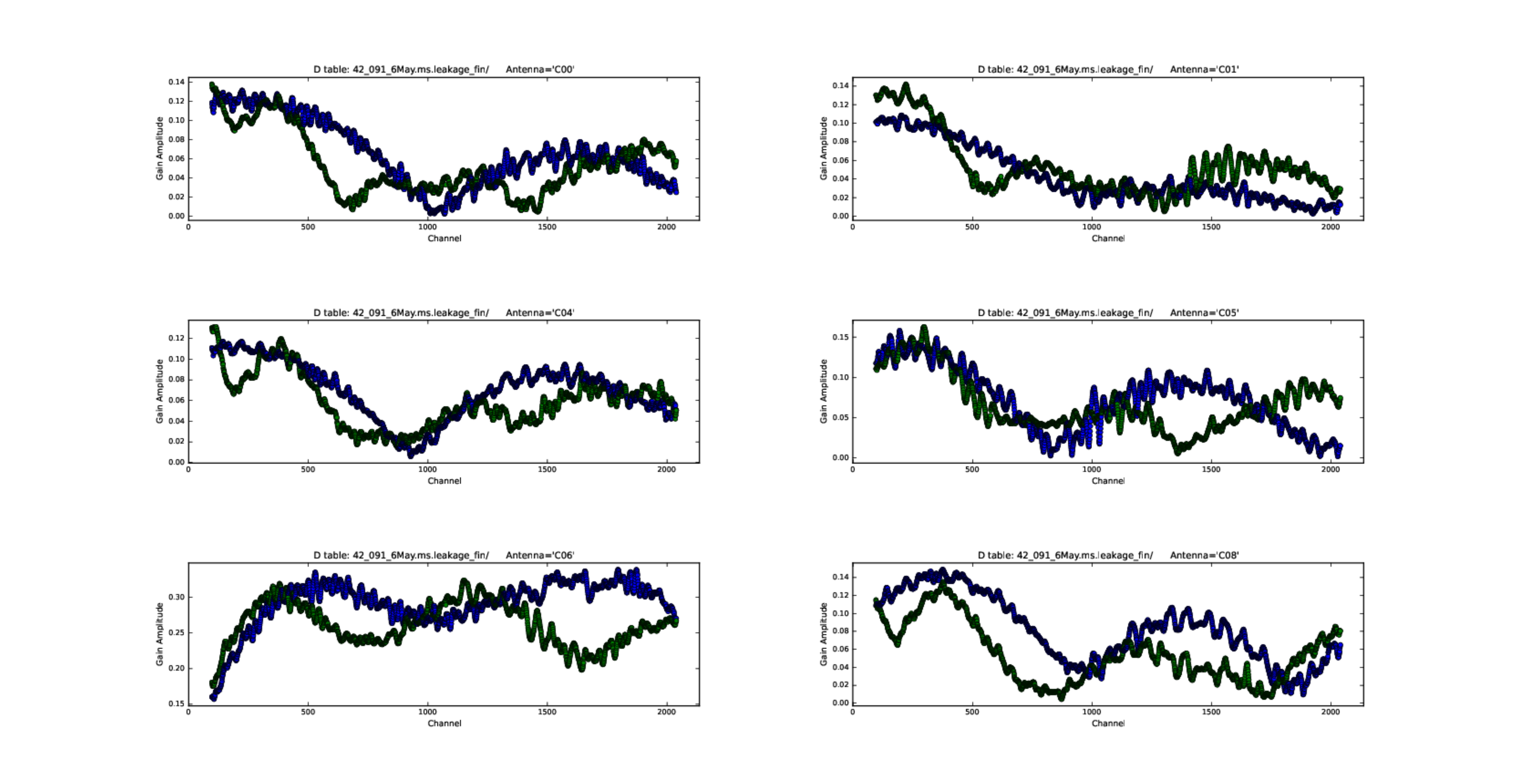}
\includegraphics[width=19cm,trim=185 40 20 20]{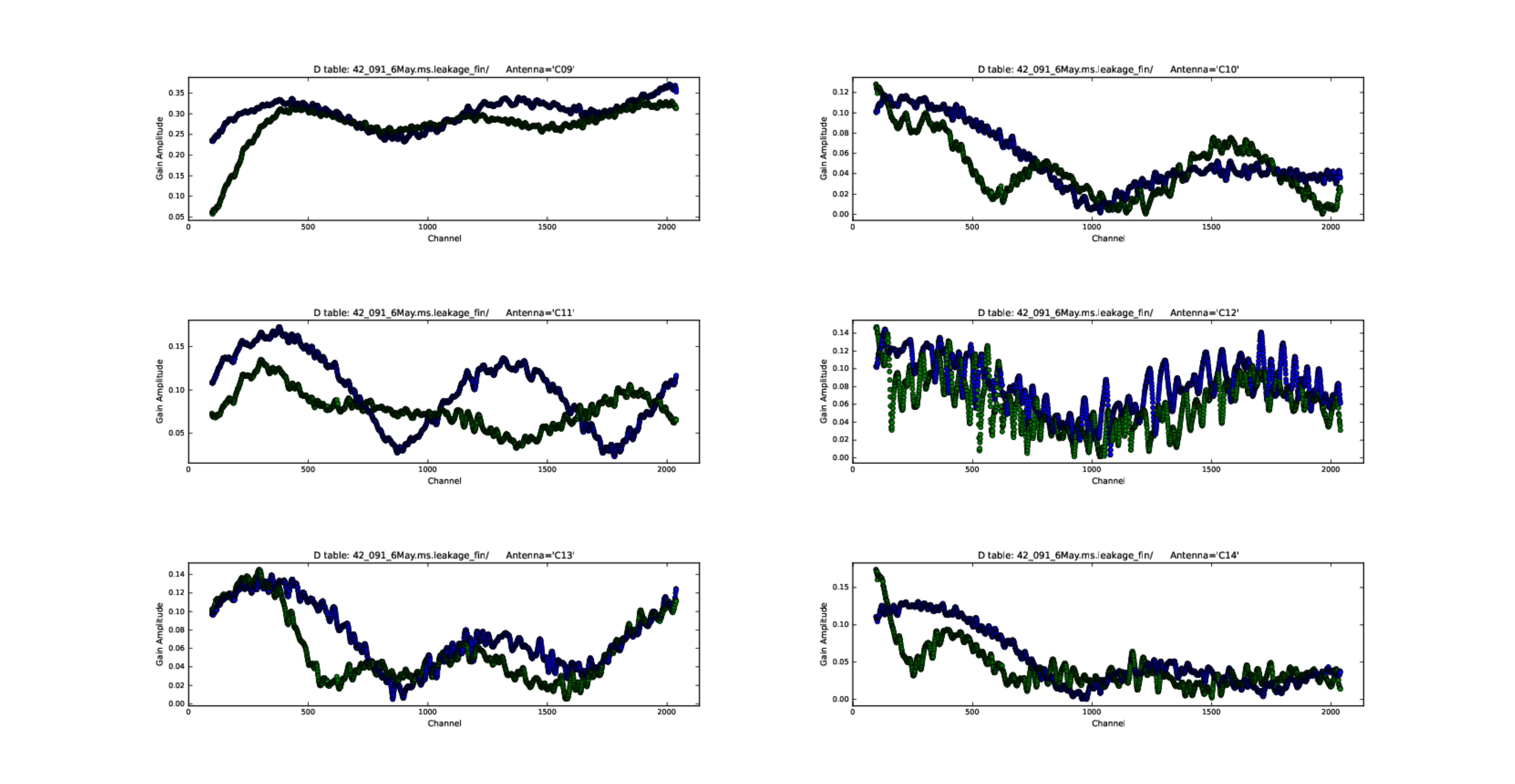}
\caption{\small Antenna `D-term' amplitudes vs channels for the Project 42\_091 from 6 May 2022 using the polarized calibrator 3C286.}
\end{figure}
\begin{figure}
\includegraphics[width=19cm,trim=185 40 20 20]{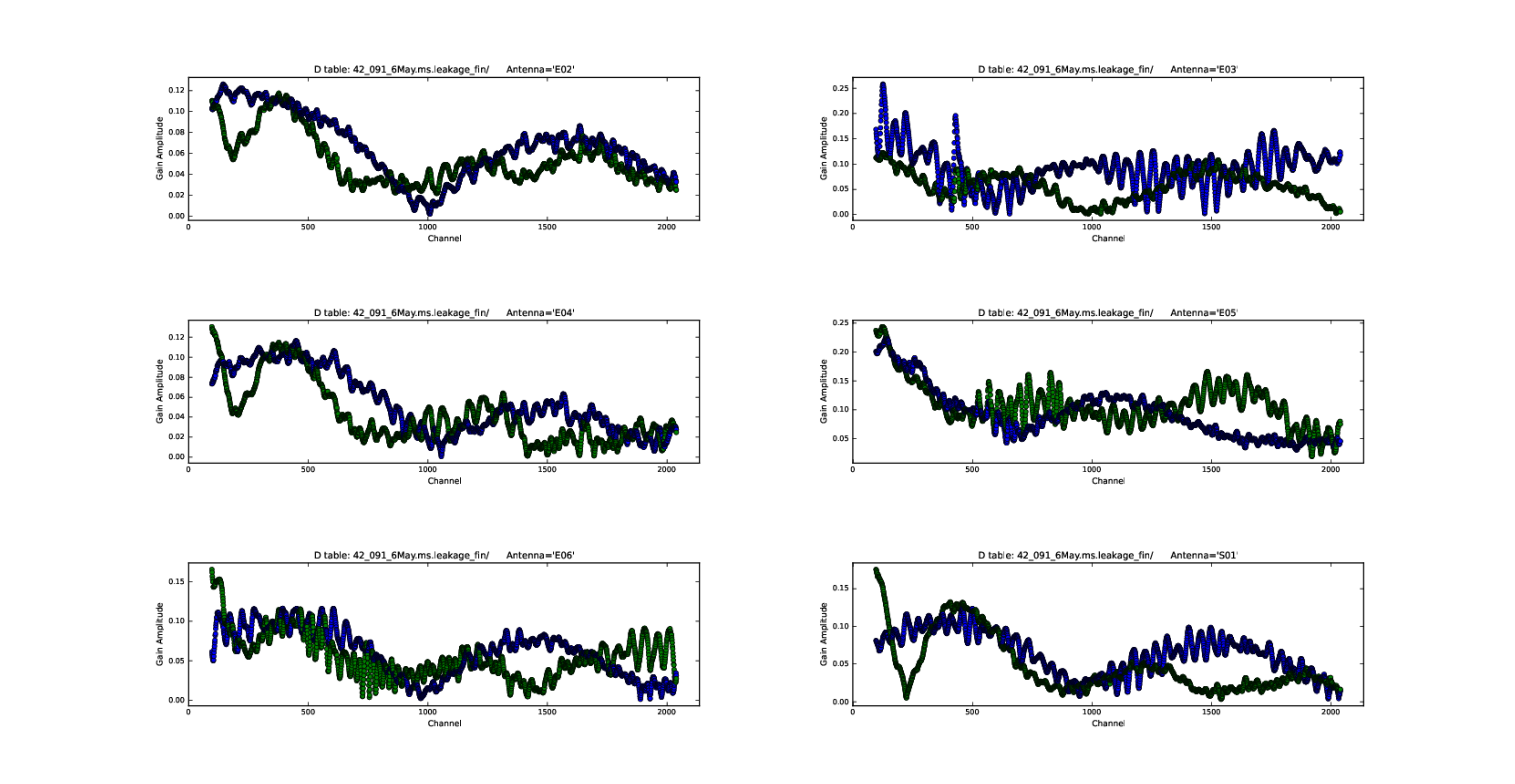}
\includegraphics[width=19cm,trim=185 40 20 20]{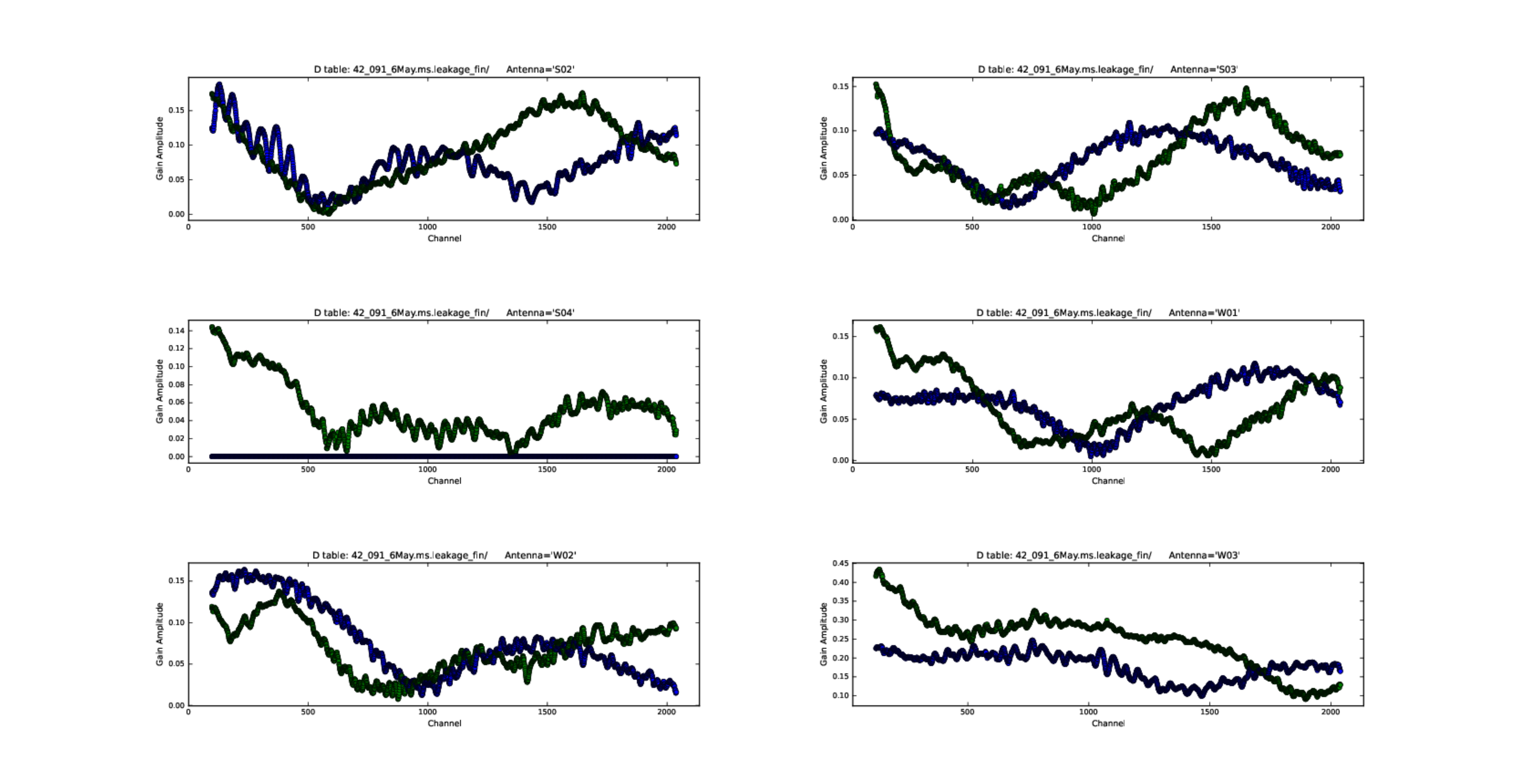}
\caption{\small Antenna `D-term' amplitudes vs channels for the Project 42\_091 from 6 May 2022 using the polarized calibrator 3C286.}
\end{figure}
\begin{figure}
\includegraphics[width=19cm,trim=185 40 20 20]{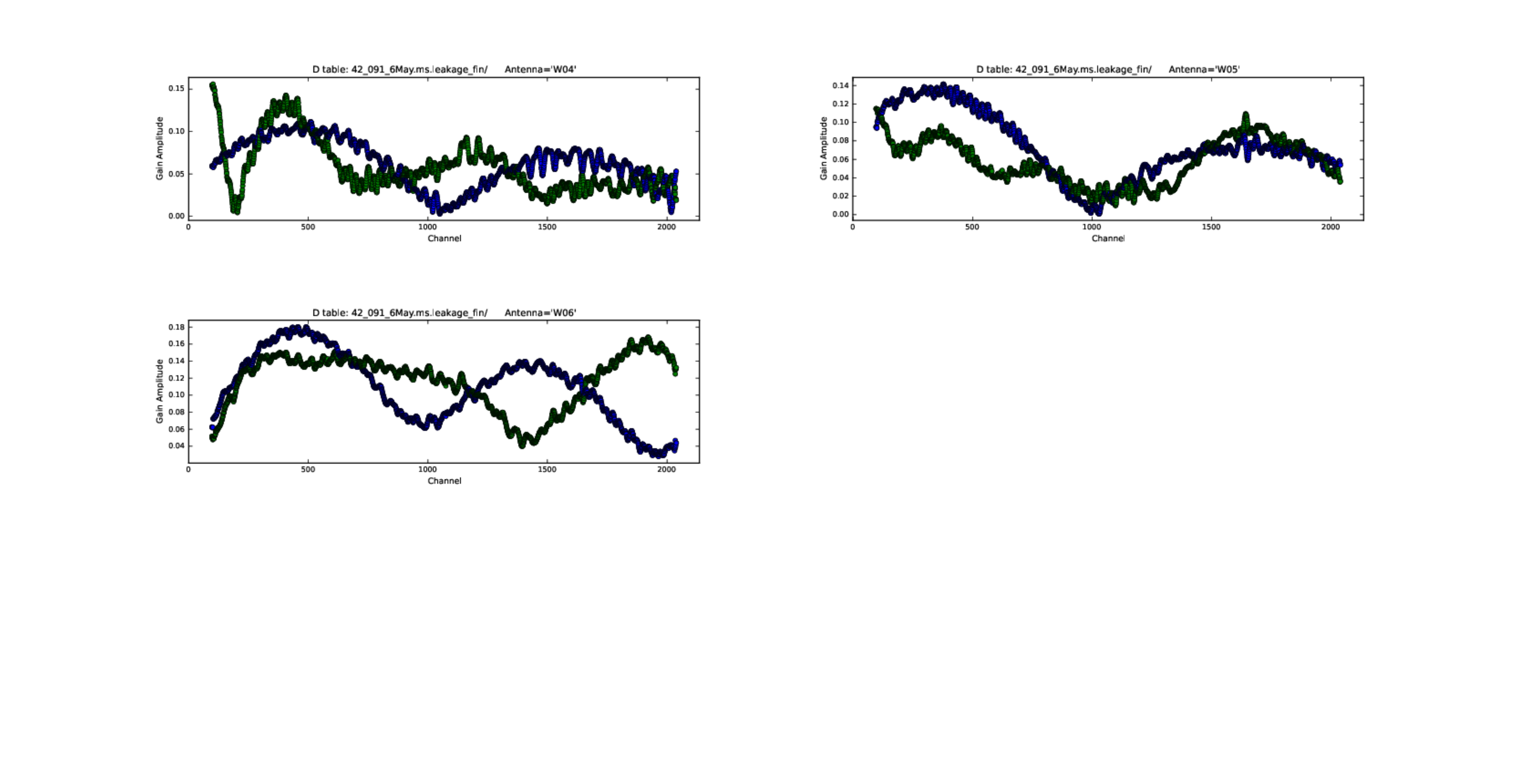}
\caption{\small Antenna `D-term' amplitudes vs channels for the Project 42\_091 from 6 May 2022 using the polarized calibrator 3C286.}
\end{figure}

\begin{figure}
\includegraphics[width=19cm,trim=185 40 20 20]{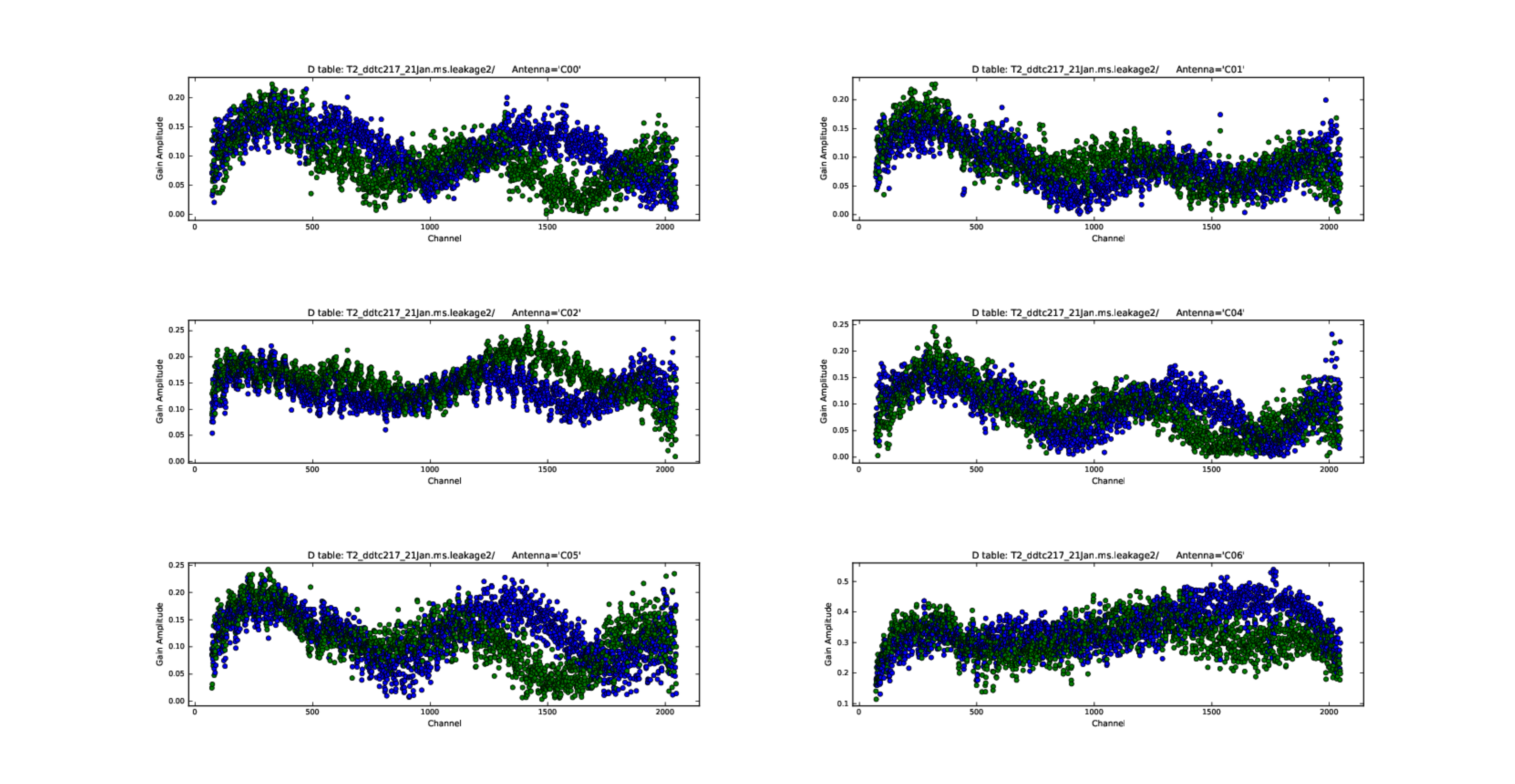}
\includegraphics[width=19cm,trim=185 40 20 20]{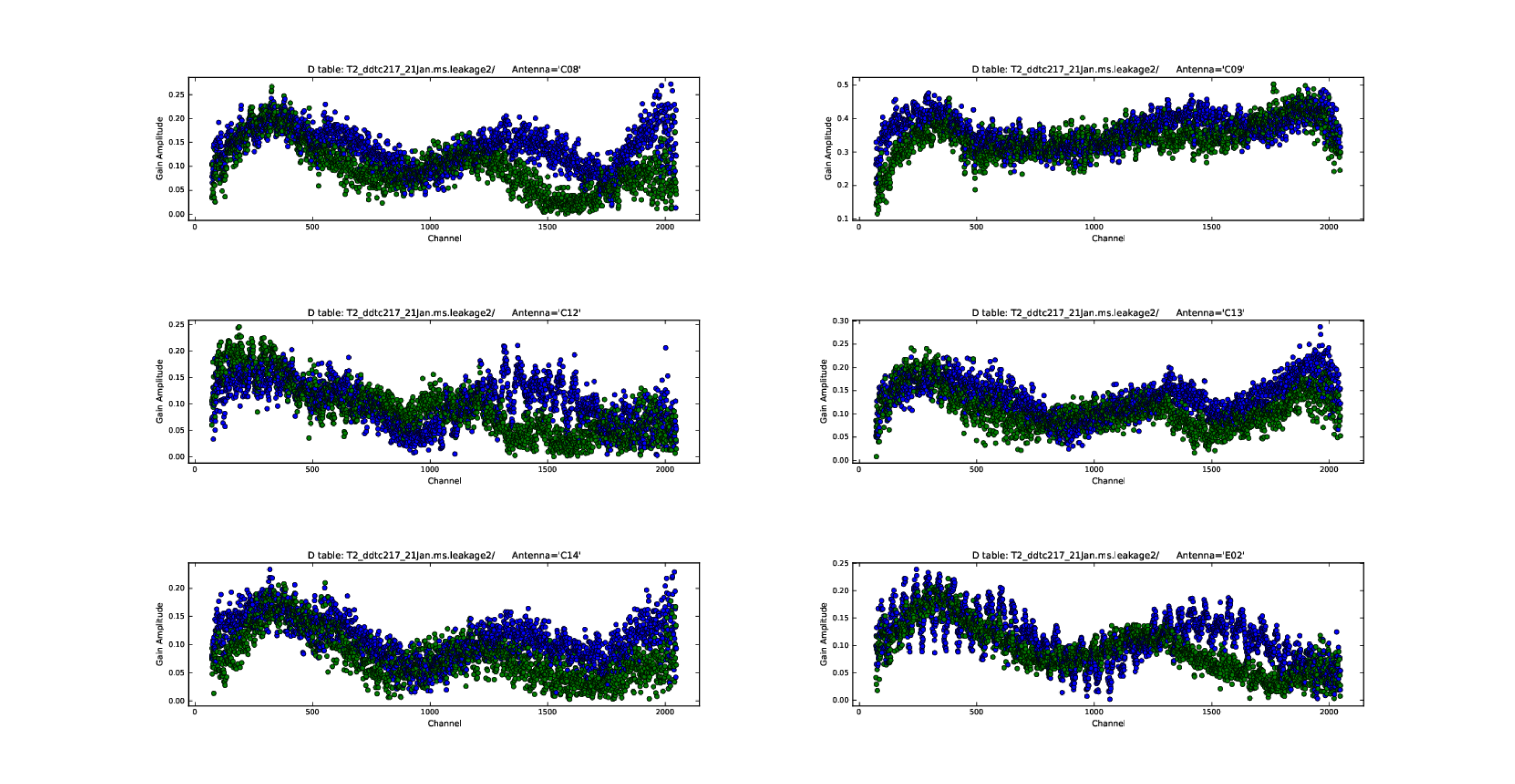}
\caption{\small Antenna `D-term' amplitudes vs channels for the Project 42\_091 from 21 Jan 2022 using the unpolarized calibrator J0713+4349.}
\end{figure}
\begin{figure}
\includegraphics[width=19cm,trim=185 40 20 20]{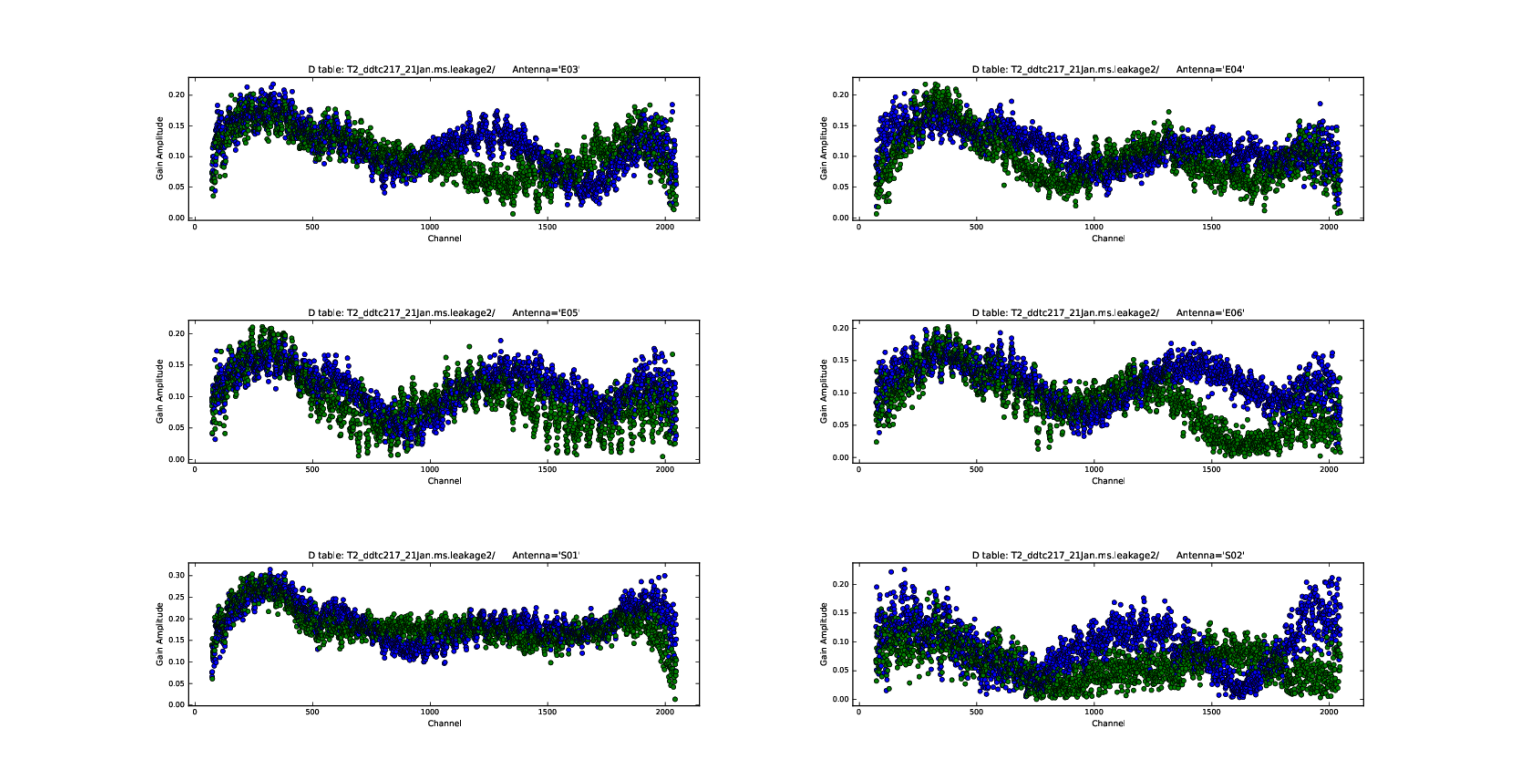}
\includegraphics[width=19cm,trim=185 40 20 20]{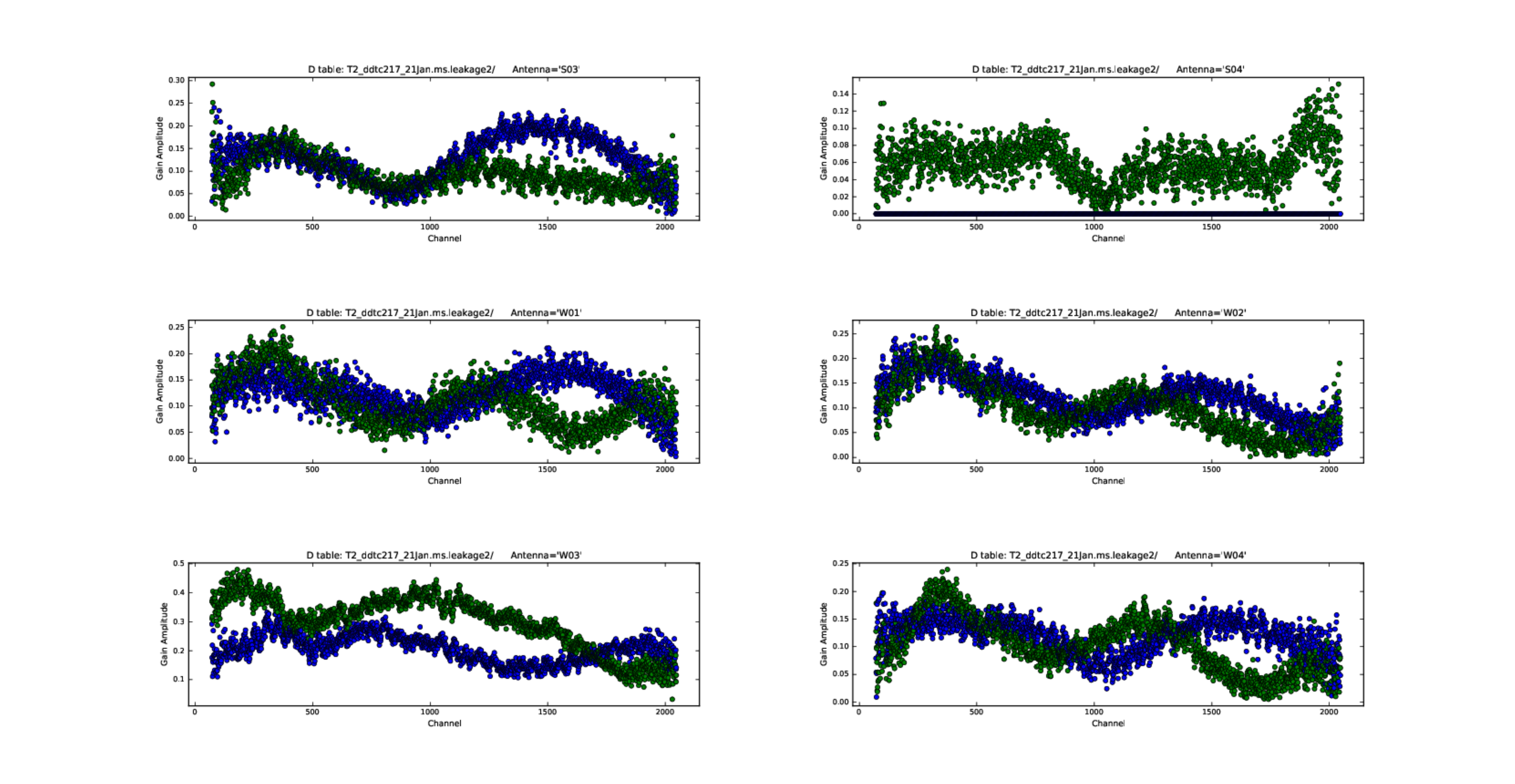}
\caption{\small Antenna `D-term' amplitudes vs channels for the Project 42\_091 from 21 Jan 2022 using the unpolarized calibrator J0713+4349.}
\end{figure}
\begin{figure}
\includegraphics[width=19cm,trim=185 400 20 20]{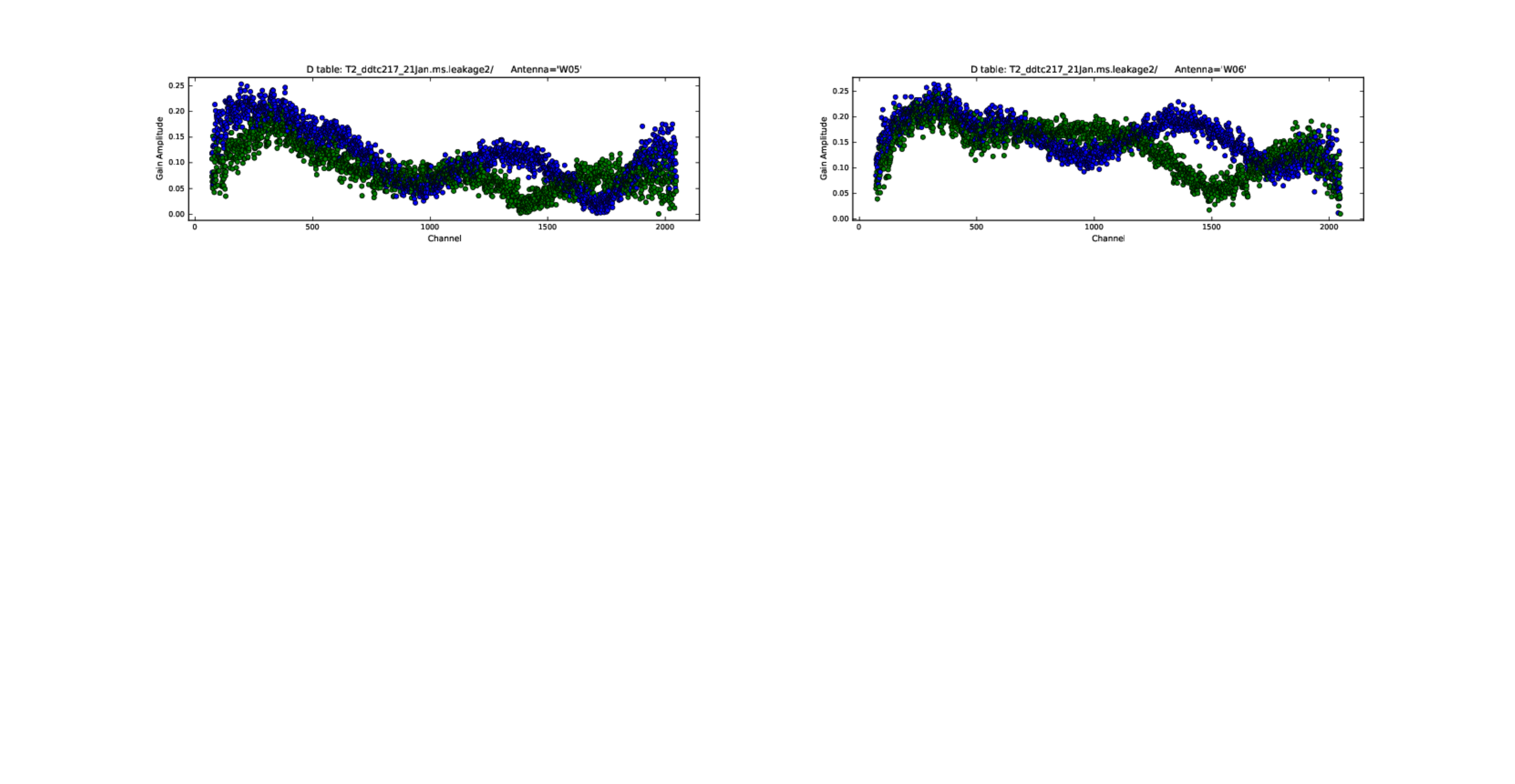}
\caption{\small Antenna `D-term' amplitudes vs channels for the Project 42\_091 from 21 Jan 2022 using the unpolarized calibrator J0713+4349.}
\end{figure}

\bibliographystyle{aasjournal}
\bibliography{References}
\end{document}